\documentclass{article}

\usepackage{spconf,amsmath,epsfig}
\usepackage{hyperref}
\usepackage{amsmath}
\usepackage{cite}
\usepackage{amssymb}
\usepackage{algorithm}
\usepackage{algpseudocode}
\usepackage{balance}
\usepackage{bbm}
\usepackage{hyperref}
\hypersetup{
    bookmarks=true,         
    unicode=false,          
    pdftoolbar=true,        
    pdfmenubar=true,        
    pdffitwindow=false,     
    pdfstartview={FitH},    
    pdftitle={CR},    
    pdfauthor={Camps-Valls},     
    pdfsubject={CR},   
    pdfcreator={Camps-Valls},   
    pdfproducer={Camps-Valls}, 
    pdfkeywords={keyword1} {key2} {key3}, 
    pdfnewwindow=true,      
    colorlinks=true,       
    linkcolor=blue,          
    citecolor=blue,        
    filecolor=blue,      
    urlcolor=blue           
}

\def\x{{\mathbf x}}

\def\C{{\mathbf C}}
\def\X{{\mathbf X}}

\def\Real{{\mathbbm{R}}}

\newcommand{\pphi}{\boldsymbol{\phi}}
\newcommand{\ppsi}{\boldsymbol{\psi}}

\newcommand{\PHI}{\boldsymbol{\Phi}}
\newcommand{\PSI}{\boldsymbol{\Psi}}

\newcommand{\y}{{\mathbf y}}

\newcommand{\W}{{\mathbf W}}
\newcommand{\Y}{{\mathbf Y}}

\usepackage{color}

\title{Consistent regression of biophysical parameters with kernel methods}

\name{Emiliano D\'iaz, Adri\'an P\'erez-Suay, Valero Laparra, Gustau Camps-Valls
	\thanks{This work was funded by the European Research Council (ERC) under the ERC-CoG-2014 SEDAL project (grant agreement 647423).}
}
\address{Image Processing Lab (IPL), Universitat de Val\`encia, Val\`encia, Spain}

\begin{document}

\maketitle

\begin{abstract}
This paper introduces a novel statistical regression framework that allows the incorporation of consistency constraints. A linear and nonlinear (kernel-based) formulation are introduced, and both imply closed-form analytical solutions. The models exploit all the information from a set of drivers while being maximally independent of a set of auxiliary, protected variables. We successfully illustrate the performance in the estimation of chlorophyll content. 
\end{abstract}


\section{Introduction}
\label{sec:intro}

Recent years have witnessed a successful adoption of statistical methods for model inversion, emulation and bio-geophysical parameter retrieval~\cite{CampsValls11mc}. Machine learning algorithms are flexible non-parametric models that fit the observations using large heterogeneous data. Machine learning models for parameter retrieval avoid complicated assumptions, provide fast and accurate estimates, and learn the complex relations directly from data.

Current operational vegetation products, like leaf area index (LAI), are typically produced with neural networks, Gross Primary Production (GPP) --as the largest global CO$_2$ flux driving several ecosystem functions-- is estimated using ensembles of random forests, kernel methods and neural networks~\cite{Tramontana16bg}, biomass has been estimated with stepwise multiple regression~\cite{Sarker11}, partial least squares regression is used for mapping canopy nitrogen~\cite{Coops03,Townsend03}, support vector regression~\cite{Smola2004} showed high efficiency in modelling LAI, fCOVER and evapotranspiration~\cite{Durbha07,Yang06}, and kernel methods in general~\cite{ShaweTaylor04,CampsValls09wiley}, 
and Gaussian Processes (GPs) in particular~\cite{Rasmussen2006}, recently provided excellent results in chlorophyll content estimation among other vegetation parameters~\cite{Furfaro06,CampsValls16grsm}.


There is however an important issue that is often disregarded: statistical models learn input-output mappings from data but very often do not respect the most elemental rules of physics. They often come up with accurate, yet inconsistent predictions. For example, bio-geo-physical parameters are typically estimated with individual, indepdent models which are typically trained separately. This common approach ignores the (potentially nonlinear) cross-relations among variables. 
Constraining the estimation problem is known in machine learning as {\em structured-output learning}, and is tightly related to {\em multitask learning}~\cite{Leiva2012,Tuia11structured}. Extension of such models to the regression setting is far from trivial, as the number of constraints increases cubically with the number of samples and outputs. Furthermore, including constraints in the regression models goes beyond consistency of model outputs; one could be interested in preserving some particular characteristics in the predictions, e.g. being independent of some ancillary information, faithful to certain variable ranges, or disregarding some information from the input (spectral) bands, just to name a few. Our notion of {\em consistency} is broad: we posit that a prediction is fully consistent with respect to some sensitive features if and only if the model's predictions are statistically independent of them.

In this work, we introduce a novel statistical regression framework that allows one to incorporate such broad consistency constraints. The framework builds upon \cite{kamishima2013independence,PerezSuay17ecmlpkdd} to minimize a functional that tries to jointly minimize the empirical error and maximize the dependence of the predictions with respect to an external subset of predictors, observations or ancillary information here called {\em sensitive} features. Two models are derived: a linear and a nonlinear, kernel-based, consistent regression model. The new {\em consistency term} trades-off accuracy for consistency, and translates into an extra regularization term that can be easily interpreted. Interestingly, the proposed models come in closed-form analytical solutions, and are very easy to implement, involving only matrix inversions.

The remainder of the paper is organized as follows. Section~\ref{sec:method} introduces notation and reviews the consistent regression models proposed. Section~\ref{sec:exp} gives experimental evidence of performance in 
chlorophyll content estimation. 
Finally, Section~\ref{sec:conc} concludes the paper with some summarizing remarks.

\section{Proposed methodology}
\label{sec:method}

This section starts by defining the notation and the concept of consistent regression.  The proposed framework for performing consistent regression learning based on cross-covariance operators for dependence estimation in Hilbert spaces is then introduced. 

\subsection{Notation and the regularization framework}

We are given $n$ samples of a response (or target) data matrix $\Y\in\Real^{n\times c}$, and $d+q$ prediction variables: $d$ driver variables $\X\in\Real^{n\times d}$ and $q$ sensitive ${\mathbf S}\in\Real^{n\times q}$. The goal is to obtain a generic prediction function (or model) $f$ for the target variable $\Y$ from the input data, $(\X,{\mathbf S})$. The goal in our framework of {\em consistent learning} is to predict $\Y$ while being maximally independent of ${\mathbf S}$.

A prediction is said to be totally consistent with respect to the sensitive features ${\mathbf S}$ if and only if $\widehat{\Y} \perp {\mathbf S}$. Therefore, two main ingredients are needed to perform consistent predictions: we need to ensure independence of the predictions on the sensitive variables, and simultaneously to obtain a good approximation of the target variables. 

The proposed function $f$ tries to learn the relation between observed input-output data pairs $(\x_1,\y_1)$, $\ldots,$ $(\x_n,\y_n)$ $\in{\mathcal X}\times {\mathcal Y}$ such that it generalizes well (good predictions $\hat{\y}_\ast=f(\x_\ast)\in{\mathcal Y}$ for the unseen input data point $\x_\ast\in{\mathcal X}$), and the predictions should be as independent as possible of the sensitive features (variables, auxiliary information or even observations). As such, the following functional should be optimized: 
\begin{equation}\label{eq:framework}
{\mathcal L} = \dfrac{1}{n}\sum_{i=1}^n V(f(\x_i),\y_i) 
                + \lambda~\Omega(\|f\|_{\mathcal H}) 
                + \mu~I(f(\x),{\bf s}),
\end{equation}
where $V$ is the error cost function, $\Omega(\|f\|_{\mathcal H})$ acts as a regularizer of the predictive function and controls the smoothness and complexity of the model, and $I(f(\x),{\bf s})$ measures the independence between the model's predictions and the protected variables. Note that one aims to minimize the amount of information that the model shares with the sensitive variables while controlling the trade-off between fitting and independence through hyperparameters $\lambda$ and $\mu$. By setting $\mu=0$ one obtains the ordinary (Tikhonov's regularized) functional, and by setting $\lambda=0$ one obtains the unregularized versions of this framework.

The framework admits many variants depending on the cost function $V$, regularizer $\Omega$ and the independence measure, $I$. For example, in~\cite{Kamishima:2012}, the function $f$ was the logistic regression classifier and $I$ was a simplification of the mutual information estimate. Despite the good results reported in~\cite{Kamishima:2012}, these choices do not allow one to solve the problem in closed-form, nor to cope with more than one sensitive variable at a time, since the proposed mutual information is an uni-dimensional dependence measure. In the following section, we elaborate on this framework by using the concept of cross-covariance operators in Hilbert spaces, which lead to closed-form solutions and permit one to deal with several sensitive variables simultaneously.

\subsection{Consistent Linear Regression}

Let us now provide a straightforward instantiation of the proposed framework for consistent linear regression (CLR). 
We will adopt a linear predictive model for $f$, i.e. the matrix of predictions for a test data matrix $\X_\ast$ is given by $\hat \Y_\ast = \X_\ast\W$, the mean square error for the cost function $V = \|\Y-\X\W\|_2^2$ and the standard $\ell_2$ regularization for model weights $\Omega:= \|\W\|_2^2$. Other choices could be made, leading to alternative formulations. In order to measure dependence, we will rely on the cross-covariance operator between the predictions and the sensitive variables in Hilbert space. Let us consider two spaces ${\mathcal Y}\subseteq \Real^{c}$ and ${\mathcal S}\subseteq \Real^{q}$, where random variables $(\hat\y,{\bf s})$ are sampled from the joint distribution ${\mathbb P}_{\y{\bf s}}$. Given a set of pairs ${\mathcal D}=\{(\hat\y_1,{\bf s}_1),\ldots,(\hat\y_n,{\bf s}_n)\}$ of size $n$ drawn from ${\mathbb P}_{\y{\bf s}}$, an empirical estimator of HSIC~\cite{Gretton05} allows us to define
$$I:=\text{HSIC}({\mathcal Y},{\mathcal S},{\mathbb P}_{\y{\bf s}}) = \|\C_{ys}\|_{\text{HS}}^2 = \frac{1}{n^2}\text{Tr}(\tilde\Y^\top \tilde{\mathbf S} \tilde{\mathbf S}^\top\tilde\Y),$$
where $\|\cdot\|_{\text{HS}}$ is the Hilbert-Schmidt norm, $\C_{ys}$ is the empirical cross-covariance matrix between predictions and sensitive variables\footnote{The covariance matrix is ${\mathcal C}_{\y{\bf s}} = {\mathbb E}_{\y{\bf s}}(\y{\bf s}^\top) - {\mathbb E}_{\y}(\y){\mathbb E}_{{\bf s}}({\bf s}^\top)$, where ${\mathbb E}_{\y{\bf s}}$ is the expectation with respect to ${\mathbb P}_{\y{\bf s}}$, and ${\mathbb E}_{\y}$ is the marginal expectation with respect to  ${\mathbb P}_{\y}$ (hereafter we assume that all these quantities exist).}, $\tilde\Y$ and $\tilde{\bf S}$ represent the feature-centered $\Y$ and ${\bf S}$ respectively, and $\text{Tr}(\cdot)$ denotes the trace operation. We want to stress that HSIC allows us to estimate dependencies between multidimensional variables, and that HSIC is zero if an only if there is no second-order dependence between $\hat\y$ and ${\bf s}$. In the next section we extend the formulation to higher-order dependencies with the use of kernels~\cite{Scholkopf02,ShaweTaylor04}. 

Plugging these definitions of $f$, $V$, $\Omega$ and $I$ in Eq.~\eqref{eq:framework}, one can easily show that the solution has the following closed-form solution for weight estimates
\begin{equation}\label{eq:disp_lin}
\widehat\W = (\tilde\X^\top \tilde\X + \lambda~{\bf I} + \frac{\mu}{n^2}~\tilde\X^\top \tilde{\mathbf S} \tilde{\mathbf S}^\top \tilde\X)^{-1} \tilde\X^\top \Y, \end{equation}
where consistency is trivially controlled with $\mu$, which acts as an additional regularization term. Also note that when $\mu=0$ the ordinary (regularized) least squares solution is obtained.


\begin{figure*}[t!]
\small
\centerline{
\begin{tabular}{ccc}
\includegraphics[width=5.6cm]{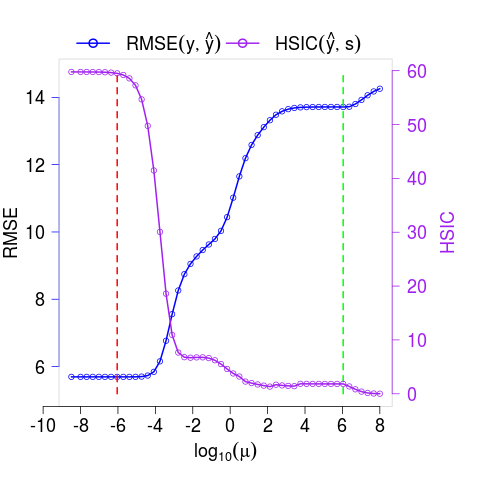}&
\includegraphics[width=5.6cm]{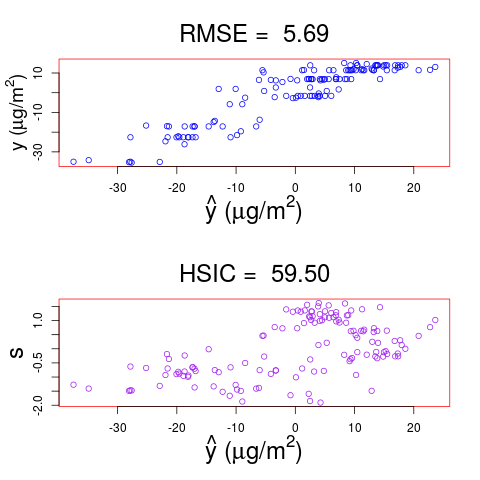}&
\includegraphics[width=5.6cm]{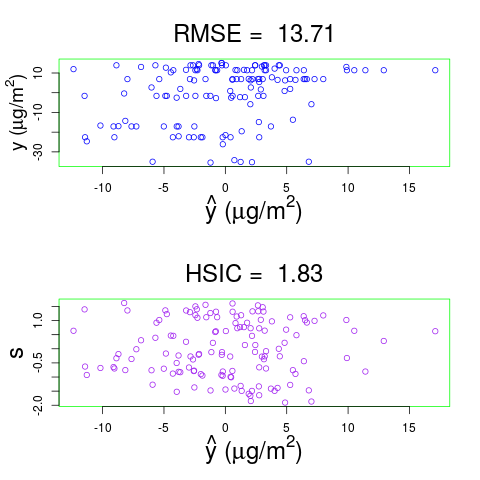}\\
\includegraphics[width=5.6cm]{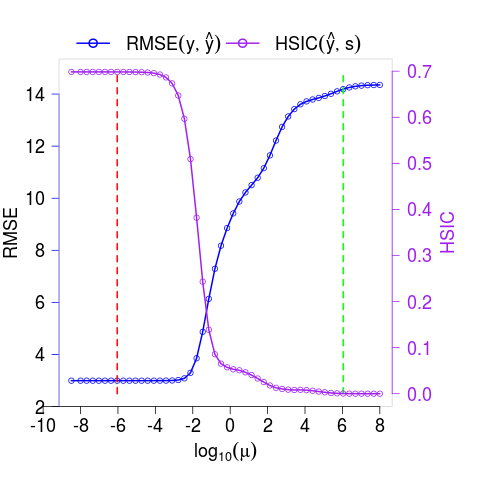}&
\includegraphics[width=5.6cm]{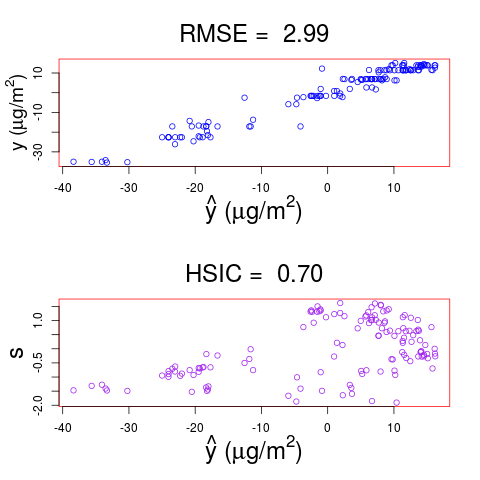}&
\includegraphics[width=5.6cm]{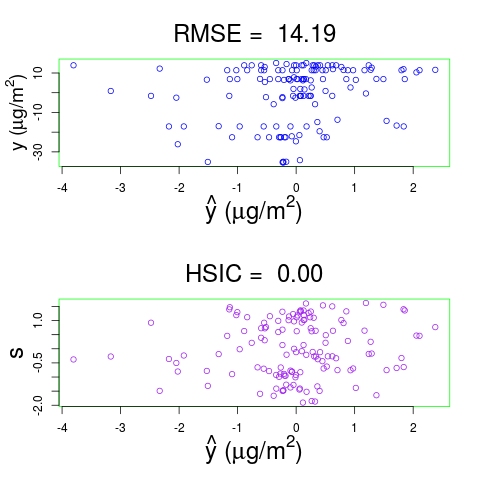}\\
\end{tabular}
}
\caption{Evolution of the RMSE [$\mu$g/m$^2$] and HSIC for predicting Chl-a with either linear (top) or kernel (bottom) regression as a function of the consistency parameter $\mu$. Scatter plots illustrate the accuracy (blue points, predicted vs. observed chlorophyll content) and consistency (purple points, predicted chlorophyll content vs. sensitive band) for two choices of $\mu$ (low and high consistency correspond to red and green respectively).}
\label{consistency} 
\end{figure*}

\subsection{Consistent Kernel Regression}

Let us now extend the previous model to the nonlinear case in terms of the prediction function, the regularizer and the dependence measure by means of reproducing kernels~\cite{Scholkopf02,ShaweTaylor04}. We call this method the consistent kernel regression (CKR) model. We proceed in the standard way in kernel machines by mapping data $\X$ and ${\mathbf S}$ to a Hilbert space ${\mathcal H}$ via the mapping functions $\pphi(\cdot)$ and $\ppsi(\cdot)$ respectively. This yields $\PHI,\PSI\in{\mathcal H}\subseteq\Real^{d_{\mathcal H}}$, where $d_{\mathcal H}$ is the (unknown and possibly infinite) dimensionality of mapped points in ${\mathcal H}$. The corresponding kernel matrices can be defined as: $\tilde{\bf K}=\tilde\PHI\tilde\PHI^\top$ and $\tilde{\bf K}_S=\tilde\PSI\tilde\PSI^\top$. 
Now the prediction function is $\hat \Y = \PHI\W_{\mathcal H}$, the regularizer is $\Omega:= \|\W_{\mathcal H}\|_2^2$, and the dependence measure $I$ is the HSIC estimate between predictions $\hat \Y$ and sensitive variables ${\mathbf S}$, which can now be estimated in Hilbert spaces: 
$I:=\text{HSIC}({\mathcal Y},{\mathcal H},{\mathbb P}_{\y{\bf s}}) = \|\C_{ys}\|_{\text{HS}}^2$.  
Now, by plugging all these terms in the cost function, using the representer's theorem $\W_{\mathcal H} = \tilde{\PHI}^\top\boldsymbol{\Lambda}$ and after some simple linear algebra, 
we obtain the dual weights in closed-form
\begin{equation}\label{eq:disp_kern}
	\boldsymbol{\Lambda} = (\tilde{\bf K} + \lambda {\bf I} + \frac{\mu}{n^2} \tilde{\bf K}_S \tilde{\bf K})^{-1} \Y,
\end{equation}
which can be used for prediction with a new point $\x_\ast$ by using $\hat\y_\ast = {\bf k}_\ast\boldsymbol{\Lambda}$, where ${\bf k}_\ast = [K(\x_\ast,\x_1),\ldots,K(\x_\ast,\x_n)]^\top$. Note that in the case where $\mu = 0$ the method reduces to standard kernel ridge regression (KRR) method~\cite{ShaweTaylor04}. Note that centering points in feature spaces can be done implicitly with kernels~\cite{ShaweTaylor04}: a kernel matrix ${\bf K}$ is centered by doing $\tilde{\bf K} = {\bf H}{\bf K}{\bf H}$, where ${\bf H}= {\bf I} - \frac{1}{n}\mathbbm{1}\mathbbm{1}^\top$.

\section{Experiments}
\label{sec:exp}

This section presents the results of the application of our consistent regression framework to a remote sensing 
problem.
In particular, we illustrate the performance of the proposed method to retrieve consistent estimates of chlorophyll content from hyperspectral images. 

\subsection{Data collection}

The data were obtained in the SPARC-2003 (SPectra bARrax Campaign) and SPARC-2004 campaigns in Barrax, Spain. The region consists of approximately 65\% dry land and 35\% irrigated land. The methodology applied to obtain the {\em in situ} leaf-level Chl$_{ab}$ data consisted of measuring samples with a calibrated CCM-200 Chlorophyll Content Meter in the field. 
Chl measurements were between 2 and 55 $\mu$g/cm$^2$. 
Additionally, 30 random bare soil spectra with zero chlorophyll 
value were added to broaden the dataset to non-vegetated samples. 
Concurrently, we used CHRIS images Mode 1 (62 spectral bands, 34m spatial resolution at nadir). The images were preprocessed, geometrically and atmospherically corrected. A total of $n=136$ datapoints in a $62$-dimensional space and the measured chlorophyll concentration constitute the database. 

\subsection{Results}

The experiment deals with the prediction of 
chlorophyll content while forcing the model to be as independent as possible from the bands beyond the NIR. It is physically understood that the chlorophyll content drives reflectance mostly in the red edge. 
Hence, our sensitive variables are the channels far beyond the NIR spectrum. 
Figure \ref{consistency} shows the results that were obtained. We show the root mean square error (RMSE) and the HSIC for different levels of the consistency parameter $\mu$ (and, implicitly, the corresponding optimal $\lambda(\mu)$) for both the linear and non-linear models. 

Several conclusions can be obtained. First, one can readily recognize, in both cases, a trade-off between obtaining accurate predictions and imposing consistency. As we increase the consistency parameter $\mu$ we obtain predictions with greater independence to the sensitive variables but the accuracy of the prediction deteriorates. 
Second, although more accurate predictions can be obtained with the non-linear, kernel model (RMSE=2.99 mg/m$^3$ versus RMSE=5.69 mg/m$^3$), the rate of deterioration of the accuracy when consistency is imposed is greater for the non-linear, kernel model than for its linear counterpart.  Figure \ref{consistency} also illustrates the quality of predictions and consistency for two choices of the consistency parameter $\mu$. It can be noted that good models in terms of accuracy (red point) lead to more correlated predictions with the sensitive bands, while enforcing the constraints (blue point) lead to higher independence but poor fitting results for both linear (RMSE=13.71 mg/m$^3$) or nonlinear (RMSE=14.19 mg/m$^3$) models.

\section{Conclusions}\label{sec:conc}

This paper presented a novel statistical regression framework that allows one to incorporate consistency constraints. The methodology confers, to both linear and nonlinear statistical regression, methods whose solution can be expressed in closed-form. The models exploit all the information from a set of covariates while being maximally independent of a set of auxiliary, protected variables. We successfully illustrated the performance for the estimation of  chlorophyll content while being independent to particular spectral bands.  

\small
\bibliographystyle{IEEEbib}
\bibliography{extra,bib_total,gprbib_clean,biblio,DSPKM,gpc,gus}

\begin{thebibliography}{10}

\bibitem{CampsValls11mc}
G.~Camps-Valls, D.~Tuia, L.~G\'omez-Chova, and J.~Malo, Eds.,
\newblock {\em Remote Sensing Image Processing},
\newblock Morgan \& Claypool, Sept 2011.

\bibitem{Tramontana16bg}
G.~Tramontana, M.~Jung, G.~Camps-Valls, K.~Ichii, B.~Raduly, M.~Reichstein,
  C.~R. Schwalm, M.~A. Arain, A.~Cescatti, G.~Kiely, L.~Merbold,
  P.~Serrano-Ortiz, S.~Sickert, S.~Wolf, and D.~Papale,
\newblock ``Predicting carbon dioxide and energy fluxes across global fluxnet
  sites with regression algorithms,''
\newblock {\em Biogeosciences Discussions}, vol. 2016, pp. 1--33, 2016.

\bibitem{Sarker11}
L.~R. Sarker and J.~E. Nichol,
\newblock ``{Improved forest biomass estimates using ALOS AVNIR-2 texture
  indices},''
\newblock {\em Rem. Sens. Env.}, vol. 115, no. 4, pp. 968--977, 2011.

\bibitem{Coops03}
N.~C. Coops, M-L. Smith, M.E. Martin, and S.~V. Ollinger,
\newblock ``Prediction of eucalypt foliage nitrogen content from
  satellite-derived hyperspectral data,''
\newblock {\em IEEE Trans. Geosc. Rem. Sens.}, vol. 41, no. 6, pp. 1338--1346,
  Jun 2003.

\bibitem{Townsend03}
P.A. Townsend, J.R. Foster, R.A.~Jr. Chastain, and W.S. Currie,
\newblock ``Application of imaging spectroscopy to mapping canopy nitrogen in
  the forests of the central {A}ppalachian {M}ountains using {H}yperion and
  {AVIRIS},''
\newblock {\em IEEE Trans. Geosc. Rem. Sens.}, vol. 41, no. 6, pp. 1347--1354,
  June 2003.

\bibitem{Smola2004}
A.~J. Smola and B.~Sch{\"o}lkopf,
\newblock ``A tutorial on support vector regression,''
\newblock {\em Statistics and Computing}, vol. 14, pp. 199--222, 2004.

\bibitem{Durbha07}
S.S. Durbha, R.L. King, and N.H. Younan,
\newblock ``Support vector machines regression for retrieval of leaf area index
  from multiangle imaging spectroradiometer,''
\newblock {\em Rem. Sens. Env.}, vol. 107, no. 1-2, pp. 348--361, 2007.

\bibitem{Yang06}
F.~Yang, M.A. White, A.R. Michaelis, K.~Ichii, H.~Hashimoto, P.~Votava, A-Xing
  Zhu, and R.R. Nemani,
\newblock ``{Prediction of Continental-Scale Evapotranspiration by Combining
  MODIS and AmeriFlux Data Through Support Vector Machine},''
\newblock {\em IEEE Trans. Geosc. Rem. Sens.}, vol. 44, no. 11, pp. 3452--3461,
  nov. 2006.

\bibitem{ShaweTaylor04}
J.~Shawe-Taylor and N.~Cristianini,
\newblock {\em Kernel {M}ethods for {P}attern {A}nalysis},
\newblock Cambridge University Press, 2004.

\bibitem{CampsValls09wiley}
G.~Camps-Valls and L.~Bruzzone, Eds.,
\newblock {\em Kernel methods for Remote Sensing Data Analysis},
\newblock Wiley \& Sons, UK, Dec 2009.

\bibitem{Rasmussen2006}
C.~E. Rasmussen and C.~K.~I. Williams,
\newblock {\em {G}aussian Processes for Machine Learning},
\newblock The MIT Press, New York, 2006.

\bibitem{Furfaro06}
R.~Furfaro, R.~D. Morris, A.~Kottas, M.~Taddy, and B.~D. Ganapol,
\newblock ``{A Gaussian Process Approach to Quantifying the Uncertainty of
  Vegetation Parameters from Remote Sensing Observations},''
\newblock {\em AGU Fall Meeting Abstracts}, pp. A261+, Dec 2006.

\bibitem{CampsValls16grsm}
G.~Camps-Valls, J.~Verrelst, J.~Mu\~{n}oz{-}Mar\'i, V.~Laparra,
  F.~Mateo-Jimenez, and J.~Gomez-Dans,
\newblock ``A survey on gaussian processes for earth observation data analysis:
  A comprehensive investigation,''
\newblock {\em IEEE Geoscience and Remote Sensing Magazine}, , no. 6, June
  2016.

\bibitem{Leiva2012}
J.~Leiva, L.~G\'omez-Chova, and G.~Camps-Valls,
\newblock ``Multitask remote sensing data classification,''
\newblock {\em {IEEE} Transactions on Geoscience and Remote Sensing}, vol. 50,
  Oct 2012.

\bibitem{Tuia11structured}
Devis Tuia, Jordi Mu{\~n}oz-Mar\'{\i}, Mikhail~F. Kanevski, and Gustavo
  Camps-Valls,
\newblock ``Structured output {SVM} for remote sensing image classification,''
\newblock {\em Signal Processing Systems}, vol. 65, no. 3, pp. 301--310, 2011.

\bibitem{kamishima2013independence}
Toshihiro Kamishima, Shotaro Akaho, Hideki Asoh, and Jun Sakuma,
\newblock ``The independence of fairness-aware classifiers,''
\newblock {\em 2013 IEEE 13th International Conference on Data Mining
  Workshops}, vol. 00, pp. 849--858, 2013.

\bibitem{PerezSuay17ecmlpkdd}
A.~P\'erez-Suay, V.~Laparra, G.~Mateo-Garc\'ia, J.~Mu~{ \~n }~oz Mar\'i,
  L.~G\'omez-Chova, and G.~Camps-Valls,
\newblock ``Fair kernel learning,''
\newblock in {\em European Conference on Machine Learning (ECML)}, Skopje,
  Macedonia, 18-22 September 2017.

\bibitem{Kamishima:2012}
Toshihiro Kamishima, Shotaro Akaho, Hideki Asoh, and Jun Sakuma,
\newblock {\em Fairness-Aware Classifier with Prejudice Remover Regularizer},
  pp. 35--50,
\newblock Springer Berlin Heidelberg, Berlin, Heidelberg, 2012.

\bibitem{Gretton05}
A.~Gretton, R.~Herbrich, and A.~Hyv{\"a}rinen,
\newblock ``Kernel methods for measuring independence,''
\newblock {\em Journal of Machine Learning Research}, vol. 6, pp. 2075--2129,
  2005.

\bibitem{Scholkopf02}
B.~Sch{\"o}lkopf and A.~Smola,
\newblock {\em Learning with Kernels -- {S}upport {V}ector {M}achines,
  Regularization, Optimization and Beyond},
\newblock MIT Press Series, 2002.

\end{thebibliography}

\end{document}